\documentclass[preprint]{revtex4}
\usepackage{amsmath} 
\usepackage{amsthm} 
\usepackage{amssymb}	
\usepackage{graphicx} 
\usepackage{slashed}
\usepackage{hyperref}

\newcommand{\aaf}[3]{\left< #1 \right| #2 \left| #3 \right>}
\newcommand{\asf}[3]{\left< #1 \right| #2 \left| #3 \right]}

\newcommand{\aad}[1]{\left< #1 \right>}

\newcommand{\ssd}[1]{\left[ #1 \right]}

\newcommand{\floor}[1]{\lfloor #1 \rfloor}

\begin{document}

\author{David Chester}
\email{dchester@ucla.edu}
\affiliation{Department of Physics and Astronomy,
UCLA, Los Angeles, CA 90095-1547, USA}
\title{BCJ Relations for One-Loop QCD Integral Coefficients}

\setcounter{page}{1}

\begin{abstract}
We present a set of one-loop integral coefficient relations in
QCD. The unitarity method is useful for exposing one-loop amplitudes
in terms of tree amplitudes. The coefficient relations are induced by
tree-level BCJ amplitude relations. We provide examples for box,
triangle, and bubble coefficients. These relations reduce the total
number of independent coefficients needed to calculate one-loop QCD
amplitudes.
\end{abstract}

\maketitle

\section{Introduction}

Devising efficient methods for calculating scattering amplitudes has
been useful to confirm the validity of the Standard Model. Even at
tree level, the number of Feynman diagrams dramatically increases as
the number of legs increases. For seven, eight, and nine external
gluons there are already 2485, 34300, and 559405 Feynman diagrams
needed at tree level~\cite{Mangano:1990by}.  Of course, with modern
techniques we can obtain the amplitude $\mathcal{A}_n$ without
calculating any Feynman diagrams.  At tree-level, the amplitude is
decomposed into a color-stripped partial amplitude $A_n$, which
separates the color from the kinematics.  The Parke--Taylor formula
gives simple form of maximally helicity violating (MHV) or anti-MHV
partial amplitudes~\cite{PT,ManganoParkeXu}. On shell recursion
developed by Britto, Cachazo, Feng and Witten (BCFW) can be used to
find any helicity~\cite{Britto2005499,BCFW}. To compute the total
amplitude $\mathcal{A}_n$, we need to compute the $n!$ partial
amplitudes $A_n$, corresponding to the different permutations of the
legs. However, not all $n!$ partial amplitudes are independent. Since
we have a trace over the color generators, the partial amplitudes have
cyclic symmetry, leaving $(n-1)!$ independent partial amplitudes. The
partial amplitudes also satisfy a reflection property and the U(1)
photon decoupling identity, which reduces the number of independent
partial amplitudes. (See
e.g. Refs.~\cite{Mangano:1990by,Dixon:1996wi}.)  Remarkably, there are
more tree level partial amplitude identities. The Kleiss--Kuijf
relations~\cite{KK} reduces the number of independent partial
amplitudes to $(n-2)!$. Furthermore, the Bern--Carrasco--Johansson
(BCJ) amplitude relations~\cite{Bern:2008qj, BjerrumBohr:2009rd,
  Stieberger:2009hq, Yin} give $(n-3)!$ independent partial
amplitudes. This paper applies these ideas to reduce the number of
independent integral coefficients at one loop.

At one-loop, we consider on-shell diagrams instead of Feynman
diagrams. We apply the unitary method finding the value of the loop
amplitude with the loop momentum on-shell. Furthermore, when we apply
the unitarity cuts, one-loop amplitudes with massless external legs
can be reconstructed in terms of products of tree amplitudes.  The
coefficients of basis integrals are fully determined from
four-dimensional tree amplitudes~\cite{Bern:1994zx,Bern:1994cg} and
rational remainders from $D$-dimensional
ones~\cite{BernMorgan,Bern:1996ja,Anastasiou:2006jv,Giele:2008ve,Badger:2008cm}. For
a modern review of on-shell and unitarity methods for one-loop QCD
amplitudes, we refer the reader to Ref.~\cite{Ita:2011hi}.

Since tree amplitudes determine the integral coefficient within the
unitarity approach, we expect that the integral coefficients satisfy
similar identities as the tree amplitudes themselves. In particular,
we show that the tree-level BCJ amplitude relations can be used to
derive integral coefficient identities. Since the loop momenta always
have two on-shell solutions, we have to decompose the integral
coefficient into two pieces. It is these pieces which actually satisfy
the coefficient relations, rather than the total coefficient.

We demonstrate that tree level identities significantly decrease the
total number of independent integral coefficients. These relations
could be used to either improve the efficiency of one-loop amplitude
calculations or to provide a stability or other cross checks for the
integral coefficients (e.g. see Ref.~\cite{Berger:2008sj}).

This paper is organized as follows: To demonstrate that the one-loop
integral coefficients satisfy BCJ integral coefficient relations, we
start by reviewing the tree-level BCJ relations and the unitarity
method in section~\ref{sec:2}. In section~\ref{sec:3}, we derive the
general expressions for the one-loop BCJ integral coefficient
relations from the tree-level BCJ amplitude relations. In
section~\ref{sec:4}, we explicitly provide examples and confirm that
the BCJ integral coefficient relation is satisfied. Finally, we
conclude and briefly discuss possible future work in section~\ref{sec:5}.

\section{From Trees to Loops}
\label{sec:2}

\subsection{Introduction to Tree-Level BCJ Amplitude Relations}

In this section, we review the unitarity method and the necessary
tree-level identities needed.  In the next section, we will show that
the BCJ amplitude relations can be used with the unitarity method to
find new relations between integral coefficients. Therefore, we start
by reviewing the BCJ amplitude relations for future reference.

 We focus on color-stripped partial amplitude. The
 color-stripped partial amplitude, $A_n$, is related to the full
 amplitude via~\cite{Dixon:1996wi}
\begin{equation}
\mathcal{A}^{\textrm{tree}}_n = g^{n-2}\sum_{\sigma\in S_n/Z_n} 
\textrm{Tr}(T^{a_{\sigma(1)}}T^{a_{\sigma(2)}}\dots 
T^{a_{\sigma(n)}})A_n^{\textrm{tree}}(\sigma(1),\sigma(2),\dots, \sigma(n)),
\end{equation}
where $S_n/Z_n$ represents the $n!$ permutations of external legs
divided by the $n$ cyclic permutations which are removed because they
would give the same color trace. Partial amplitudes only depend on
kinematic variables, and the labels of the partial amplitude signify
the momenta and helicities of the particles.

The BCJ amplitude relations at tree level are connected to
color-kinematics duality~\cite{Bern:2008qj}.  The duality forces the
Jacobi identity on the kinematic numerators, which naturally provide
tree-level partial amplitude relations beyond what is contained in the
Kleiss--Kuijf relations. The BCJ relations imply that only $(n-3)!$
of the partial amplitudes are independent. A convenient choice is to fix the first 3
legs. At four and five points, the BCJ tree amplitude relations are
\begin{eqnarray}
A_4^{\textrm{tree}}(1,2,\{4\},3) &=& 
A_4^{\textrm{tree}}(1,2,3,4)\frac{s_{14}}{s_{24}}, \nonumber \\
\label{eqn:BCJ}
A_5^{\textrm{tree}}(1,2,\{4\},3,5) &=& \frac{A_5^{\textrm{tree}}(1,2,3,4,5)
(s_{14}+s_{45})+A_5^{\textrm{tree}}(1,2,3,5,4)s_{14}}{s_{24}}, \nonumber \\
A_5^{\textrm{tree}}(1,2,\{4,5\},3) &=& \frac{-A_5^{\textrm{tree}}(1,2,3,4,5)s_{34}s_{15}
-A_5^{\textrm{tree}}(1,2,3,5,4)s_{14}(s_{245}+s_{35})}{s_{24}s_{245}}.
\end{eqnarray}
The original BCJ paper also includes general $n$-point formulas for
generating the BCJ relations. Notice how the BCJ amplitude relations
fixes the first three legs for the set of independent partial
amplitudes.

To express amplitudes, we will use the spinor-helicity formalism,
which gives remarkably compact expressions for certain tree
amplitudes. Tree amplitudes with zero or one plus/minus helicities are
zero (except for the three-point amplitudes). The $n$-point
Parke--Taylor formula~\cite{PT,ManganoParkeXu} expresses the MHV
partial amplitudes in a remarkably simple manner,
\begin{equation}
A_n^{\textrm{tree}}(1^+,\dots,i^-,\dots,j^-,\dots,n^+) = i 
 \frac{\aad{ij}^4}{\aad{12}\aad{23}\dots\aad{n1}}.
\end{equation}
To find all other helicity configurations, BCFW recursion, for
example, can be applied to calculate any amplitude from the
Parke--Taylor formula~\cite{Britto2005499}.  (See, for example,
Ref.~\cite{Truijen} for an exhaustive review with examples.)

\subsection{Unitary Cuts and One-Loop Integral Basis Coefficients}
We now review the unitarity method, which is used find one-loop
amplitudes as a linear combination of integral coefficients times
basis integrals. At one loop any massless amplitude can be decomposed
into a set of basis integrals consisting of scalar boxes, triangles,
and bubbles, plus rational terms for
QCD~\cite{Melrose:1965kb,vanNeerven:1983vr,Bern:1993kr}.  This reduces
the problem of calculating one-loop amplitudes to determining a set
rational coefficients. 

Britto, Cachazo, and Feng showed how generalized unitarity could be
used to find the box coefficients~\cite{Britto:2004nc}. The work of
Ossola, Papadopolous, and Pittau~\cite{Ossola:2006us} and
Forde~\cite{Forde:2007mi} extends this to the triangle and bubble
coefficients.  Unitarity cuts can be used
to recycle tree amplitudes into the loop level integral
coefficients. Therefore, we can generate all one-loop amplitudes from
tree amplitudes; the coefficients are simply products of tree
amplitudes with loop momenta on-shell.

\begin{figure}
\begin{center}
\includegraphics[scale=.6]{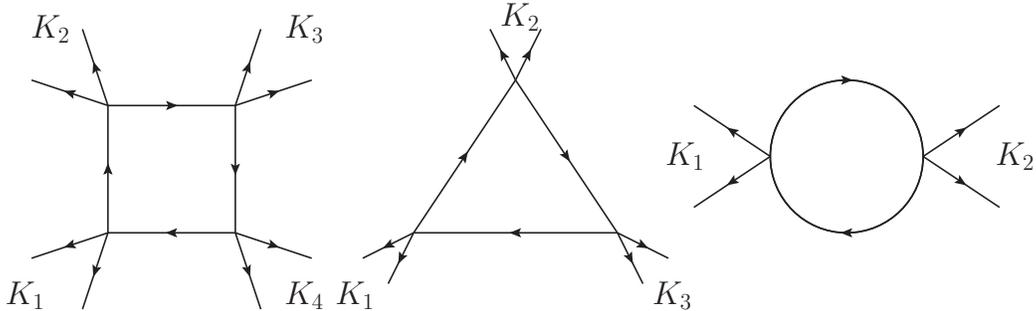}
\caption{The box, triangle, and bubble cuts. At each corner there are an arbitrary of external lines.}
\label{btb}
\end{center}
\end{figure}

We will study the boxes, triangles, and bubbles necessary for massless
QCD in the following subsections. In particular, we are interested in
the total number of cuts needed as well as the loop momentum solutions
for such cuts. In order to apply unitarity cuts to a diagram, we put
the loop momenta on shell which is imposed on all internal lines in Fig.~\ref{btb}.

To calculate the full one-loop amplitude, each possible unitary cut
must be evaluated. Using generalized unitarity this allows us to
determine the coefficients of basis integrals in terms of which the
amplitude is expressed,
\begin{equation}
A = \sum_{i=1}^{n_{b}} c_i \textrm{Int}_i + R\,,
\end{equation}
where $i$ runs over the total number of basis integrals $n_b$,
Int$_i$ is a scalar integral, and $R$ represents the rational terms which will be neglected throughout this work.

\subsubsection{Box Cuts}

\begin{figure}
\begin{center}
\includegraphics[scale=.6]{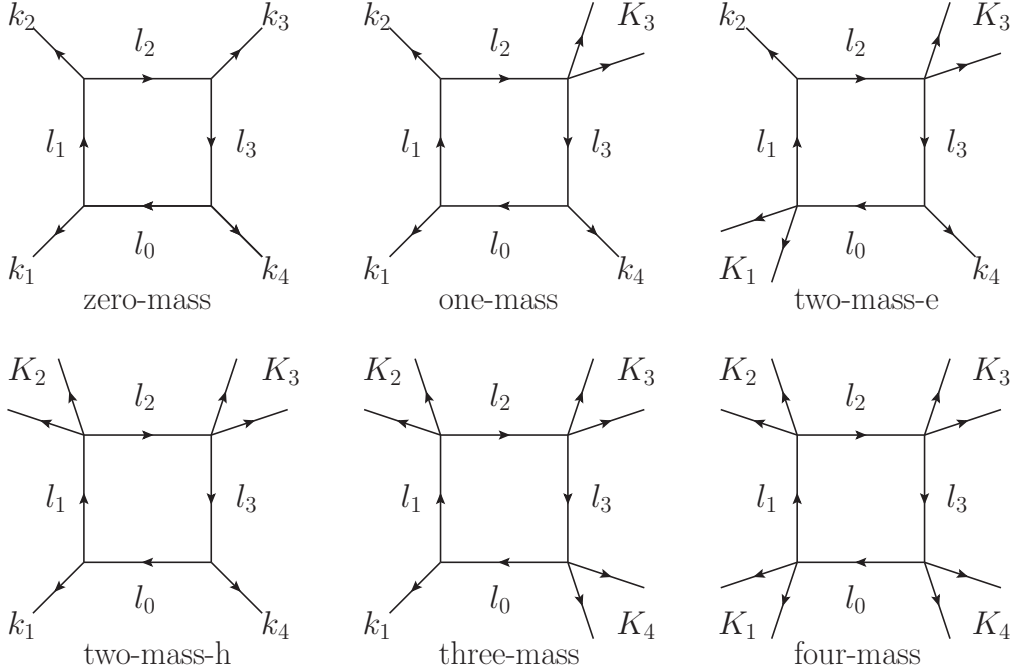}
\caption{The zero-mass, one-mass, two-mass-e, two-mass-h, three-mass,
  and four-mass box cuts are shown above. Two-mass-e and two-mass-h
  stand for `easy' and `hard'.  The $K_i$ are sums of massless momenta
  and the $k_i$ is the momentum of a single external massless leg.}
\label{fig1}
\end{center}
\end{figure}

To start our study of QCD amplitudes, we focus on the box coefficients
which can be determined from the box cuts shown in Fig.~\ref{btb}.
Next, we classify the different types of box cuts.  There are
zero-mass, one-mass, two-mass-e, two-mass-h, three-mass, and four-mass
box cuts, which is shown in Fig.~\ref{fig1}. Only the four-point
amplitudes have a zero-mass box. There are $n$ $n$-point one-mass
boxes for $n\ge5$, since there are $n$ ways that the $n-3$ particles
could be put together in the corner. For two-mass-h boxes, there are
$(n-5)n$ ways with $n\ge 6$. The number of two-mass-e boxes would be
the same, but we are now overcounting by a factor of two due to the
symmetry of the two single legs being across from each
other. Therefore, there are $(n-5)n/2$ two-mass-h boxes with $n\ge 6$.
When $n\ge7$, there are $n{n-5\choose 2}$ three-mass boxes. For
four-mass boxes, there are three contributing boxes. If $n\ge8$ and
$n\!\mod4 =0$, then there are $n/4$ contributing boxes. This
corresponds to when all four corners of the box cut have the same
number of legs, such as the only four-mass eight-point cut would
have. If $n\ge9$ and $n$ is even, then there are an additional
$n/2$. Finally, if $n\ge9$, then there are
$\sum_{i=1}^{n-8}n{\lfloor(i+1)/2\rfloor+1\choose2}$ additional
four-mass boxes.

The unitarity cuts put the loop momenta on-shell, which gives a
quadratic equation with two solutions. If there is at least one
massless leg on the cut, we utilize the following compact solution~\cite{Berger:2008sj,Risager:2008yz}, 
\begin{eqnarray}
l_0^{\pm,\mu} = \frac{\aaf{1^\mp}{\slashed{K}_2\slashed{K}_3\slashed{K}_4\gamma^\mu}{1^\pm}}{2\aaf{1^\mp}{\slashed{K}_2\slashed{K}_4}{1^\pm}},\textrm{ }\textrm{ }\textrm{ }\textrm{ }\textrm{ } l_1^{\pm,\mu} = -\frac{\aaf{1^\mp}{\gamma^\mu\slashed{K}_2\slashed{K}_3\slashed{K}_4}{1^\pm}}{2\aaf{1^\mp}{\slashed{K}_2\slashed{K}_4}{1^\pm}},\nonumber \\
l_2^{\pm,\mu} = \frac{\aaf{1^\mp}{\slashed{K}_2\gamma^\mu\slashed{K}_3\slashed{K}_4}{1^\pm}}{2\aaf{1^\mp}{\slashed{K}_2\slashed{K}_4}{1^\pm}},\textrm{ }\textrm{ }\textrm{ }\textrm{ }\textrm{ } l_3^{\pm,\mu} = -\frac{\aaf{1^\mp}{\slashed{K}_2\slashed{K}_3\gamma^\mu\slashed{K}_4}{1^\pm}}{2\aaf{1^\mp}{\slashed{K}_2\slashed{K}_4}{1^\pm}}.
\label{eqn:18}
\end{eqnarray}
Note that the `$-$' solution is simply the complex conjugate of the
`$+$' solution. If we have a four-mass box, then we must resort to
using the more lengthy solution provided by Britto, Cachazo, and
Feng~\cite{Britto:2004nc}. (See Ref.~\cite{Davies:2011vt} for loop
momentum solutions in $d=4-2\epsilon$.)

To obtain an arbitrary box coefficient, we simply apply a unitarity
cut and multiply the four corresponding tree amplitudes together:
\begin{eqnarray}
A_n^{\textrm{one-loop}} \Bigr|_{\rm boxes} &=& \sum_{i=1}^m d_i \textrm{Box}_i =
 \sum_{i=1}^m (d_i^+ + d_i^-) \textrm{Box}_i, \nonumber \\
d^\pm_{i;box} &=& \frac{1}{2}A^{\textrm{tree}\pm}_{1,i} A^{\textrm{tree}\pm}_{2,i} 
A^{\textrm{tree}\pm}_{3,i} A^{\textrm{tree}\pm}_{4,i},
\end{eqnarray}
where the $\pm$ on the coefficients refers to the two loop solutions,
$A^{\textrm{tree}\pm}_{j,i}$ are the tree amplitudes from the cuts,
and $m$ is the total number of boxes. Note that the true integral
coefficient $d_{i;box}$ is the sum of the two coefficients
$d_{i;box}^\pm$, but we must keep these separate to find the
coefficient relations. 

\subsubsection{Triangles}

\begin{figure}
\begin{center}
\includegraphics[scale=.6]{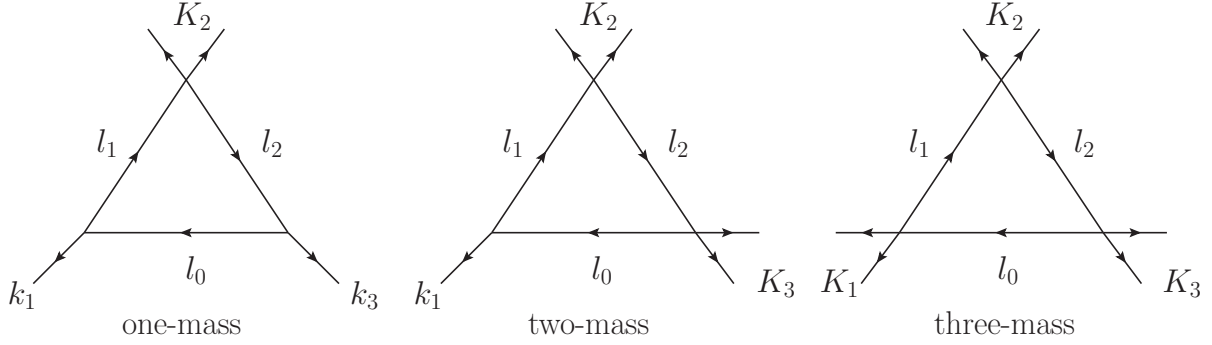}
\caption{The one-mass, two-mass, and three-mass triangle cuts are shown above.}
\label{fig2}
\end{center}
\end{figure}

Next consider the triangle integral coefficients.  We start with the
triangles by counting the number of triangle cuts. There are the
one-mass, two-mass, and three-mass triangle cuts, which are shown in
Fig.~\ref{fig2}. Once again, there are $n$ one-mass cuts. Similar to
the two-mass-e box cut, there are $(n-4)n$ two-mass triangle
cuts. Finally, for the three-mass, we have an equation similar to the
four-mass box cut, but no choose function is needed since it would
have been choose 1. If $n\mod3=0$ and $n>5$, then we have $n/3$
contributing cuts. Furthermore, if $n>6$, then we have an additional
$\sum_{i=1}^{n-5}n\lfloor\frac{i+1}{3} \rfloor$.

For finding the loop solutions for triangle diagrams, a parameter $t$
is introduced to represent the undetermined degree of freedom from
having only three unitarity cuts.  After converting Forde's loop
solution~\cite{Forde:2007mi} into our notation, we find
\begin{eqnarray}
\left< l_i^+\right| = t\left< K_1^\flat\right| + \alpha_{i1}\left< K_3^\flat\right|, &\textrm{ }\textrm{ }\textrm{ }\textrm{ }\textrm{ }& \left| l_i^+\right] = \frac{\alpha_{i2}}{t}\left| K_1^\flat\right]  + \left| K_3^\flat\right], \nonumber \\
\alpha_{01} = \frac{S_1(\gamma_{13}+S_3)}{\gamma_{13}^2-S_1 S_3},  &\textrm{ }\textrm{ }\textrm{ }\textrm{ }\textrm{ }& \alpha_{02} = -\frac{S_3(\gamma_{13}+S_1)}{\gamma_{13}^2 - S_1 S_3}, \nonumber \\
\alpha_{11} = \alpha_{01} -\frac{S_1}{\gamma_{13}}, &\textrm{ }\textrm{ }\textrm{ }\textrm{ }\textrm{ }& \alpha_{12} = \alpha_{02} - 1, \nonumber \\
\alpha_{21} = \alpha_{01} +1, &\textrm{ }\textrm{ }\textrm{ }\textrm{ }\textrm{ }& \alpha_{12} = \alpha_{02} +\frac{S_3}{\gamma_{13}},
\end{eqnarray}
where $i=0,1,2$, $S_i=K_i\cdot K_i$, and $\gamma_{13}^\pm=K_1\cdot K_3\pm\sqrt{(K_1\cdot K_3)^2-K_1^2K_3^2}$. The four-momentum representation of the loop momentum solution is
\begin{equation}
l_i^{+,\mu} = \alpha_{i2}K_1^{\flat,\mu} + \alpha_{i1}K_3^{\flat,\mu} + \frac{t}{2}\asf{K_1^\flat}{\gamma^\mu}{K_3^\flat} + \frac{\alpha_{i1}\alpha_{i2}}{2t}\asf{K_3^\flat}{\gamma^\mu}{K_1^\flat}.
\end{equation}
We can in principle consider four triangle loop solutions, since there
are two loop momenta and two gammas. However, if $S_1$ or $S_3 = 0$,
then there is only one non-zero solution for gamma. In either case, we
simply average over the two or four solutions, including the two loop
momenta. To find the second loop solution, we can take the complex
conjugate, or simply switch all of the angle brackets with the square
brackets.

Note that our $\alpha_{ij}$'s are a bit different than Forde, since we
use different conventions for the loop and external momenta. Just as
Forde showed, we find that $\alpha_{i1}\alpha_{i2} =
\alpha_{j1}\alpha_{j2}$, for $i,j = 0,1,2$. Our conventions match
those of Refs.~\cite{Berger:2008sj,Davies:2011vt}.

Following Forde's procedure to find the $t$-independent triangle coefficient we take
\begin{eqnarray}
c^\pm_{i;tri}(t) &=& \frac{1}{n_{\textrm{sol}}}A^{\textrm{tree}\pm}_{1,i} A^{\textrm{tree}\pm}_{2,i} A^{\textrm{tree}\pm}_{3,i}, \nonumber \\
\label{eqn:trisol}
c^\pm_{i;tri} &=& [\textrm{Inf}_t c^\pm_{i;tri}(t)]|_{t=0},
\end{eqnarray}
where $n_{\textrm{sol}}$ is two or four, depending on if there are one
or two independent values for gamma.  The symbol $\textrm{Inf}_t$
instructs one to Taylor expand with respect to $t$ around infinity and
keep the $t^0=1$ term to obtain the triangle integral coefficients.

\subsubsection{Bubbles}

\begin{figure}
\begin{center}
\includegraphics[scale=.6]{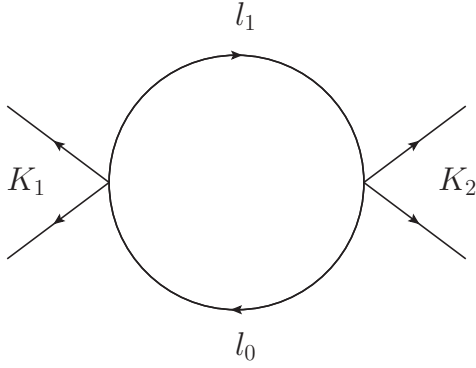}
\caption{The needed two-mass bubble cut is shown above. The one-mass bubble cuts vanish.}
\label{fig3}
\end{center}
\end{figure}

Next, we review the extraction of the bubble coefficient.
Counting the number of bubble cuts is much simpler, since there are only one-mass and two-mass diagrams. There are $n$ one-mass cuts and $\frac{n-3}{2}n$ two-mass cuts, shown in Fig.~\ref{fig3}. 
The bubble loop momentum solution has two arbitrary parameters, $t$ and $y$. Instead of using $K_2^{\flat,\mu}$, any arbitrary massless vector $\chi^\mu$ can be used. It is often convenient to choose it to be the last leg on $K_1$, which gives a simple $K_1^{\flat,\mu}$ and $\chi^\mu$. We define $K_1^{\flat,\mu}$ in terms of the massless spinor.
\begin{equation}
K_1^{\flat,\mu} = K_1^\mu -\frac{S_1}{2K_1\cdot\chi}\chi^\mu = K_1^\mu - \chi^{\flat,\mu}.
\end{equation}
Now, we are ready to express the loop solution in terms of $K_1^{\flat,\mu}$ and $\chi^{\flat\mu}$.
\begin{eqnarray}
\left<l_0^+\right| = t\left<K_1^\flat\right| + (1-y)\left< \chi^\flat\right|, &\textrm{ }\textrm{ }\textrm{ }\textrm{ }\textrm{ }& \left|l_0^+\right] = \frac{y}{t}\left|K_1^\flat\right] + \left|\chi^\flat\right], \nonumber \\ 
\left<l_1^+\right| = \left<K_1^\flat\right| - \frac{y}{t}\left< \chi^\flat\right|, &\textrm{ }\textrm{ }\textrm{ }\textrm{ }\textrm{ }& \left|l_1^+\right] = (y-1)\left|K_1^\flat\right] + t\left|\chi^\flat\right].
\end{eqnarray}
From these we can find loop momenta,
\begin{eqnarray}
l_0^{+,\mu} &=& yK_1^{\flat,\mu} + (1-y)\chi^{\flat,\mu} + \frac{t}{2}\asf{K_1^\flat}{\gamma^\mu}{\chi^\flat} + \frac{y(1-y)}{2t}\asf{\chi^\flat}{\gamma^\mu}{K_1^\flat}, \nonumber \\
l_1^{+,\mu} &=& (y-1)K_1^{\flat,\mu} - y \chi^{\flat,\mu} + \frac{t}{2}\asf{K_1^\flat}{\gamma^\mu}{\chi^\flat} + \frac{y(1-y)}{2t}\asf{\chi^\flat}{\gamma^\mu}{K_1^\flat}.
\end{eqnarray}
To find a bubble coefficient, we first find the bubble cut
contribution to the bubble coefficient, leaving in the $y$ and $t$
dependence,
\begin{equation}
b^\pm_{i;bub}(t,y) = \frac{1}{2}A^{\textrm{tree}\pm}_{1,i}(t,y) A^{\textrm{tree}\pm}_{2,i}(t,y).
\end{equation}
To obtain the full bubble coefficient independent of $y$ or $t$, one
can use the methods described in Refs.~\cite{Ossola:2006us,
  Forde:2007mi, Badger:2008cm, Davies:2011vt, Berger:2008sj}. The expression for the bubble cut
contribution to coefficient is
\begin{eqnarray}
\label{eqn:36}
b_{i;bub}^\pm &=& [\textrm{Inf}_t[\textrm{Inf}_yb_{i;bub}^\pm(t,y)]|_{y^i=Y_i}]|_{t=0}, \nonumber \\
Y_0 &=& 1,\mbox{ }\mbox{ }Y_1 = \frac{1}{2},\mbox{ }\mbox{ }Y_2 = \frac{1}{3},\mbox{ }\mbox{ }Y_3 = 
\frac{1}{4},\mbox{ }\mbox{ }Y_4 = \frac{1}{5}.
\end{eqnarray}
The bubble coefficient includes contributions from the triangles, but
we will show that these contributions also satisfy the BCJ integral
coefficient relations when we review the triangles.  We will not
review the bubble extraction in full detail, since we show that the
triangle coefficient satisfies the new coefficient relations for all
orders of $t$, implying that it hold for the total bubble coefficient
as well.

Our goal is to understand how tree-level amplitude relations can be
used to create loop-level integral coefficient relations. The BCJ
amplitude relations are needed, since the ordering of the two loop
momenta for any tree amplitude is already fixed. Since the
relations fix the third leg, we can fix one of the external legs for
any of the isolated tree amplitudes. We see that one-loop integral
coefficient relations naturally arise from the tree-level BCJ
relations. 

\section{One-Loop Amplitude Coefficient Relations}
\label{sec:3}
Now that we reviewed the calculation of integral coefficients in
arbitrary one-loop amplitudes, we focus on relations between these
coefficients. To start, we count the number of integral coefficients
needed before the BCJ integral coefficient relations are introduced.

Since the coefficient relations we will derive are independent of the
external helicities, we will typically focus on MHV amplitudes for
simplicity. Fixing the external helicity configuration, there are $n!$ 
external leg orderings, but the properties of the color
trace leave only $(n-1)!/2$ to consider. The
$(n-1)!$ factor comes from the cyclic symmetry and the factor of $1/2$
comes from the reflective symmetry of the trace over the color
generators. Therefore, one would naively expect there to be
$m(n-1)!/2$ independent integral coefficients, where
$m=m_{box}+m_{tri}+m_{bub}$ is the number of cuts made per ordering of external momenta.
In this section, we will show that the number of
independent integral coefficients and tree amplitudes is actually
smaller, since the BCJ relations can be recycled into the one-loop
level by the unitarity method.

Since each $d_i$ contains two adjacent loop momenta, the BCJ amplitude
relations may be easily used if we fix the loop momenta to be the
first two legs of the tree amplitude. Let us start by systematically
considering possible box cuts of increasing complexity.

\subsection{BCJ box integral coefficient identities}

We start by considering the simplest box cuts.
The first non-trivial example is a
five-point one-loop amplitude, which contains only one-mass box
cuts. After taking into account the cyclic and reflective symmetries
of the partial amplitudes, one would expect $(n-1)!/2=12$ independent
partial amplitudes. Furthermore, for each of these twelve amplitudes,
there are five box cuts, giving 60 coefficients to compute. We would
like to demonstrate that after taking the BCJ relations into account,
we can reduce the number of independent coefficients to 30.

\begin{figure}
\begin{center}
\includegraphics[scale=.7]{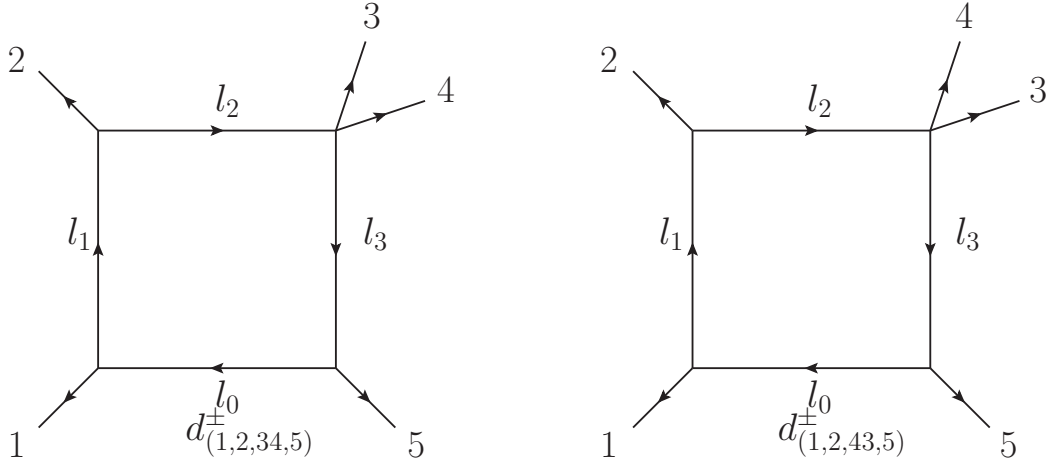}
\caption{We consider two box cuts needed, which are identical up to a twisting of the $K_3$ leg. These two coefficients have the same loop solution, which allows for a tree level BCJ amplitude relation to be used to relate the two integral coefficients.}
\label{boxex}
\end{center}
\end{figure}

Consider the following two integral coefficients $d^\pm_{(1,2,34,5)}$
and $d^\pm_{(1,2,43,5)}$, shown in Fig.~\ref{boxex}. Not only do they
have the same loop solution, but they share three tree
amplitudes. Note that the total box coefficient is found by summing
the two loop solutions, but we keep these two solutions separate to
expose the tree level BCJ amplitude relations. Let us take a closer
inspection at the dissimilar tree amplitudes containing $K_3$
\begin{eqnarray}
d^\pm_{3,(1,2,34,5)} &=& A_4^{\textrm{tree}}(l^\pm_3,-l^\pm_2,3,4), \nonumber\\
d^\pm_{3,(1,2,43,5)} &=& A_4^{\textrm{tree}}(l^\pm_3,-l^\pm_2,4,3). 
\end{eqnarray}
Since the two amplitudes have the same loop momentum solution, we can
use the four-point BCJ relation, Eq.~(\ref{eqn:BCJ}), to relate the
box coefficient. That is,
\begin{equation}
d^\pm_{1,2,43,5} = \frac{s_{l^\pm_3 4}}{s_{-l^\pm_2 4}}d^\pm_{1,2,34,5}.
\end{equation}
We look to find all possible five-point box integral coefficient
relations. After considering reflection symmetry as well as this new
"twist symmetry" for the legs located on the tree, we find that there
are 30 independent five-point coefficients instead of 60. Finding the
form of the 30 integral coefficient relations at five-points is as
trivial as finding the correct four-point tree-level BCJ amplitude
identity, so we will not review the one-mass boxes any further.

Next, consider six-point amplitudes. There are one-mass boxes and
two-mass diagrams to consider.  We start with the one-mass
diagrams. In most cases, there will be multiple one-mass diagrams
which have the same loop solution. Consider the following six
coefficients: $d^\pm_{1,2,345,6}$, $d^\pm_{1,2,354,6}$,
$d^\pm_{1,2,435,6}$, $d^\pm_{1,2,453,6}$, $d^\pm_{1,2,534,6}$, and
$d^\pm_{1,2,543,6}$. It is clear that we can use the BCJ relations to
remove the calculation of four coefficients. Using
Eq.~(\ref{eqn:BCJ}),
\begin{eqnarray}
d^\pm_{1,2,435,6} &=& d^\pm_{1,2,345,6}\frac{s_{l^\pm_34}+s_{45}}{s_{-l^\pm_24}} + d^\pm_{1,2,354,6}\frac{s_{l^\pm_34}}{s_{-l^\pm_24}}, \nonumber\\
d^\pm_{1,2,453,6} &=& -d^\pm_{1,2,345,6}\frac{s_{34}s_{l^\pm_35}}{s_{-l^\pm_24}s_{-l^\pm_245}} -d^\pm_{1,2,354,6}\frac{s_{l^\pm_34}(s_{-l^\pm_245}+s_{35})}{s_{-l^\pm_24}s_{-l^\pm_245}}, \nonumber\\
d^\pm_{1,2,534,6} &=&d^\pm_{1,2,354,6}\frac{s_{l^\pm_35}+s_{45}}{s_{-l^\pm_25}} + d^\pm_{1,2,345,6}\frac{s_{l^\pm_35}}{s_{-l^\pm_25}}, \nonumber\\
d^\pm_{1,2,543,6} &=& -d^\pm_{1,2,354,6}\frac{s_{35}s_{l^\pm_34}}{s_{-l^\pm_25}s_{-l^\pm_254}} -d^\pm_{1,2,345,6}\frac{s_{l^\pm_35}(s_{-l^\pm_254}+s_{34})}{s_{-l^\pm_25}s_{-l^\pm_254}}.
\end{eqnarray}
We see that at six-points, there are even more integral coefficient relations, since there are more cuts with the third leg of the tree amplitude fixed. For counting the number of independent box coefficients needed, it is important to note that not all of the six-point one-mass boxes have six coefficients with the same loop solution. In some cases, there will only be three independent coefficients with the same solution. 

Note that at six-points there are 360 one-mass coefficients, yet there are many fewer unique loop solutions. The BCJ relations relate a majority of the coefficients with the same loop momenta. We can see that if there are more than two six-point one-mass box coefficients with the same loop solution up to an overall minus sign to account for reflections, then those extra are dependent on two coefficients. In some cases, you may need to calculate a coefficient which is not needed, but this inconvenience decreases the number of coefficients calculated in the long run. For example, one loop solution has the following three coefficients: $d^\pm_{4,5,612,3}$, $d^\pm_{4,5,126,3}$, and $d^\pm_{4,5,261,3}$. Notice how we can use $d^\pm_{4,5,612,3}$ and $d^\pm_{4,5,621,3}$ as independent basis coefficients, even though we do not need to calculate the second coefficient. It is beneficial, since we are still only calculating two instead of three.

Next, we consider the six-point two-mass-e coefficients. These are a bit more complicated, since there are potentially two tree amplitudes $K_2$ and $K_3$ which can be different for the same loop solution. Once again, we group all of the coefficients with the same loop coefficient. At most, we could have four coefficients which have the same loop solution. For example, consider the following coefficients: $d^\pm_{1,23,45,6}$, $d^\pm_{1,23,54,6}$, $d^\pm_{1,32,45,6}$, and $d^\pm_{1,32,54,6}$. It is clear that we can relate the second and third coefficient to the first, but the fourth has two twisted corners on $K_2$ and $K_3$. Expanding the fourth coefficient in terms of tree amplitudes makes the identity more apparent:
\begin{equation}
d^\pm_{1,32,54,6} = \frac{1}{2}A_3^{\textrm{tree}}(l^\pm_1,-l^\pm,1)A_4^{\textrm{tree}}(l^\pm_2,-l^\pm_1,3,2)A_4^{\textrm{tree}}(l^\pm_3,-l^\pm_2,5,4)A_4^{\textrm{tree}}(l^\pm,-l^\pm_3,6).
\end{equation}
From expanding the coefficient in terms of tree amplitudes, we see that two four-point BCJ relations can be used to find this coefficient in terms of $d^\pm_{1,23,45,6}$. The three relations between the four coefficients mentioned above are
\begin{eqnarray}
d^\pm_{1,23,54,6} &=& d^\pm_{1,23,45,6}\frac{s_{l^\pm_35}}{s_{-l^\pm_25}}, \nonumber\\
d^\pm_{1,32,45,6} &=& d^\pm_{1,23,45,6}\frac{s_{l^\pm_23}}{s_{-l^\pm_13}}, \nonumber\\
d^\pm_{1,32,54,6} &=& d^\pm_{1,23,45,6}\frac{s_{l^\pm_35}s_{l^\pm_23}}{s_{-l^\pm_25}s_{-l^\pm_13}}.
\end{eqnarray}
Therefore, we have demonstrated that even if multiple corners of a cut have different orderings, the BCJ relations can still be used multiple times, as long as the related coefficients have the same loop solution. 

When continuing this analysis to higher-point amplitudes, we find that the simplification gets better as we increase $n$. The simplification occurs because there are more possible diagrams with the same loop solution. At seven-point, there are 12600 needed box coefficients, but only 1785 independent coefficients. For example, there are 24 permutations of $K_3$ for the coefficient $d^\pm_{1,2,3456,7}$, yet there are only six independent coefficients needed. We will not write out these 18 relations, but it is clear that the BCJ amplitude relations could be used to reduce the number of coefficients. Also, at seven-point we introduce three-mass box coefficients, which could have up to eigth coefficients with the same loop solution. For example, consider twisting $K_2$, $K_3$, and $K_4$ on the coefficient $d^\pm_{1,23,45,67}$. These eight coefficients can all be related to one coefficient, which gives seven relations. Once again, we will not write them down, but it would be easy to generate with the BCJ relations. One would have to be a bit more careful with writing down the relations for the twelve coefficients corresponding to the loop solution contained in $d^\pm_{1,23,456,7}$. The $K_3$ term gives a dependence on two coefficients, while the $K_2$ would add an overall factor of inverse propagators from Eq.~(\ref{eqn:BCJ}).

Interesting eight-point amplitudes to consider would be the four-mass and $d^\pm_{1,432,765,8}$, since that coefficient would depend on four coefficients. We will not go into deriving the identities, because it is fairly straightforward. Nothing new arises for higher-point boxes besides applying more complicated BCJ relations. 
\begin{figure}
\begin{center}
\includegraphics[scale=.8]{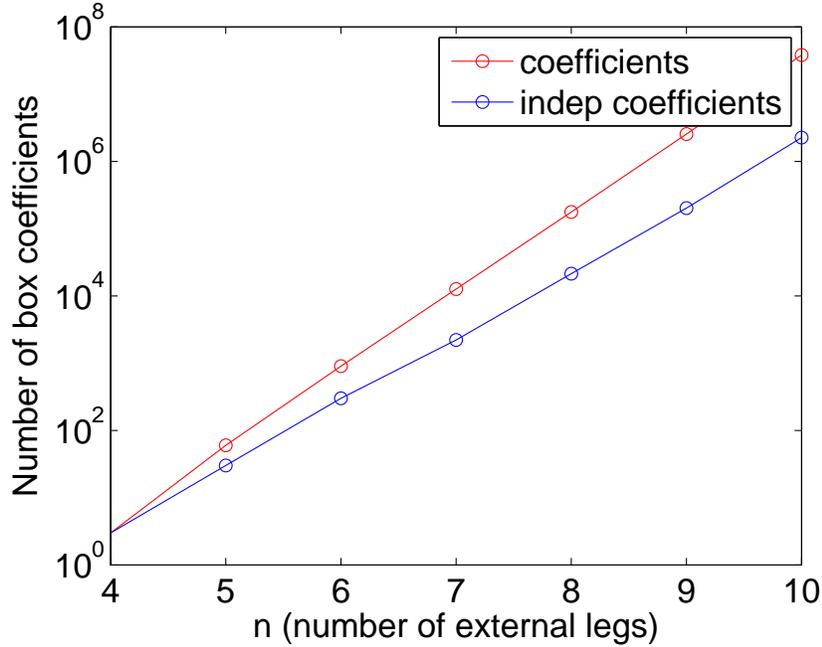}
\caption{The number of box coefficients. The red line represents the
  number of independent coefficients before using the BCJ integral
  coefficient relations, and the blue line represents the number of
  independent coefficients needed after the relations are taken into
  consideration.}
\label{plotbox}
\end{center}
\end{figure}

In Fig.~\ref{plotbox}, we plot the number of box coefficients needed before and after the BCJ integral coefficient relations have been taken into consideration. The log plot shows that as the number of external legs increases, the relations reduce a higher percentage of the coefficients. We see that by eight-points, the number of independent coefficients needed to calculate is roughly an order of magnitude less than what was naively expected. 

Finally, we would like to present a formula to count the total number
of independent coefficients, after applying the BCJ relations. The
boxes are a bit complicated, as there are different countings for the
one-mass, two-mass-e, two-mass-h, three-mass, and four-mass boxes. We
found the following expression which gave the total number of
independent coefficients $C(n)$ for $n\geq 4$.
\begin{equation}
C(n) = \sum_{i=1}^{\floor{\frac{n-4}{4}}-1} \sum_{j=1}^{\floor{\frac{n-4}{3}}-1} \sum_{k=1}^{\floor{\frac{n-4}{2}}-1} \frac{n!}{ijk(n-i-j-k)\textrm{sym}(i,j,k)},
\end{equation}
where $\floor{x}$ is the floor function and sym is a symmetry factor
which depends on whether $i$, $j$, $k$, and $n-i-j-k$, are the same or
not. Note that $i$, $j$, $k$, and $n-i-j-k$ represent the number of
external legs on each corner of the box. Naively, we would expect that
there would be $n!$ for each cut topology, but we know that the BCJ
relations allow for this new type of twist symmetry, which, up to symmetry factors,
divides the number of diagrams by the number of legs on all corners. To get the
counting exactly right, we introduced a symmetry factor sym,
\begin{equation}
\textrm{sym} = \left\{
     \begin{array}{ll}
       8 & : \textrm{ if } i=j=k=n-i-j-k \\
       2 & : \textrm{ if } i=j=k \textrm{, or any other 3 equal} \\
       1 & : \textrm{ if } i=j \textrm{ and } k=n-i-j-k\textrm{, or any other two pair equal} \\
       2/3 & : \textrm{ if } i=j\textrm{, or any other two equal} \\
       1/3 & : \textrm{ else}
     \end{array}
   \right.
\end{equation}
The symmetry factor is chosen to properly count the number of needed
diagram as well as reflection symmetry. For example, consider a
five-point box cut. There is only one type of diagram possible with
one corner having two legs and three corners with one leg. There is
only one diagram for this specification, yet there is a reflection
symmetry, making ${\rm sym} = 2$. For other cases, there could be more
diagrams needed, each with their own symmetry properties.

As we have shown, it is no surprise that the box coefficients should satisfy BCJ integral coefficient relations. Next, we investigate how similar identities can be found for the less trivial triangle coefficients.


\subsection{BCJ triangle integral coefficient identities}

Next, we study how the BCJ relations can be used to simplify triangle
integral coefficients. Exactly how the BCJ relations will come into
play is less clear, since the loop solutions contain
the parameter $t$. In particular, the inverse propagators in the BCJ
relations contain loop momenta, and therefore we will end up with
expressions for dependent triangle coefficients in terms of the
parameter $t$.

We start by considering two coefficients with the same loop solution,
say $c^\pm_{(12,34,56)}$ and $c^\pm_{(12,34,65)}$. We would like to
find a way to relate these coefficients by propagators, such that
$c_{(12,34,56)}=\frac{s_{l5}}{s_{-l_25}} c_{(12,34,56)}$, but we need
to be careful with the $t$ dependence of the amplitude and the inverse
propagators. The first natural guess would be to keep the $t$
dependence in the inverse propagators and the coefficient and apply
the proper expansion of $t$ around infinity, followed by extracting
the $t^0$ term. It turns out that this precisely works, as we confirmed
numerically. The coefficient identity is
\begin{equation}
c^\pm_{(12,34,65)}(t) =  \frac{s_{l6}(t)}{s_{-l_26}(t)}c^\pm_{(12,34,56)}(t).
\end{equation}
Finding this $t$-dependent coefficient, taking limit as $t$ goes to
infinity, and taking the $t^0$ term allows for the triangle coefficient
to be found:
\begin{equation}
c^\pm_{(12,34,65)} = [\mbox{Inf}_tc^\pm_{(12,34,65)}(t)]|_{t=0}
\end{equation}
This shows that the triangle contributions to the bubbles
should also satisfy this BCJ identity.

Now that we understand how to properly deal with the $t$ parameter,
generating triangle coefficient identities is essentially the same as
the box coefficients. We group all of the coefficients with the same
loop solution together, find the set of independent coefficients, and
write down the analogous BCJ relations needed to find the dependent
coefficients.

Now that we have demonstrated that the BCJ relations indeed hold for
triangle coefficients with Forde's analytic
approach~\cite{Forde:2007mi}, we would like to investigate how we can
use them to speed up numerical calculations. 

When considering if the BCJ relations speed up performance, there is a
caveat since the triangle coefficient only needs the zeroth order
term. However, finding $c_{(12,34,65)}$ with BCJ requires that we keep
all of the coefficients in the expansion of $t$ for $c_{(12,34,56)}$,
since the factors of inverse propagators have $t$ dependence and will
change the zeroth order dependence of the undetermined triangle
coefficient. Fortunately, all of these coefficients would be saved for
the evaluation of bubble diagrams. Therefore, the only extra
computational cost for determining $c_{(12,34,65)}$ involves finding
the coefficients from expanding $\frac{s_{l5}(t)}{s_{l_25}(t)}$ with
respect to $t$. Furthermore, it appears that all of the factors of
inverse propagators in all of the BCJ relations, even for higher than
four-points, will never have a $t^n$ term for $n>0$. Even if we did not
extract out the boxes and numerically evaluated the coefficients in a
Laurent expansion of $t$, we would only need to calculate the zeroth
and first three negative powers of the inverse propagator terms. Thus,
we have shown that the only extra computation needed the four
coefficients for $\frac{s_{l5}(t)}{s_{l_25}(t)} =
\mathcal{O}(\frac{1}{t^4}) + \sum_{i=-3}^0 a_it^i$, since
$c_{(12,34,56)}$ only goes up to powers of $t^3$ for 
Forde's method.
\begin{figure}
\begin{center}
\includegraphics[scale=.8]{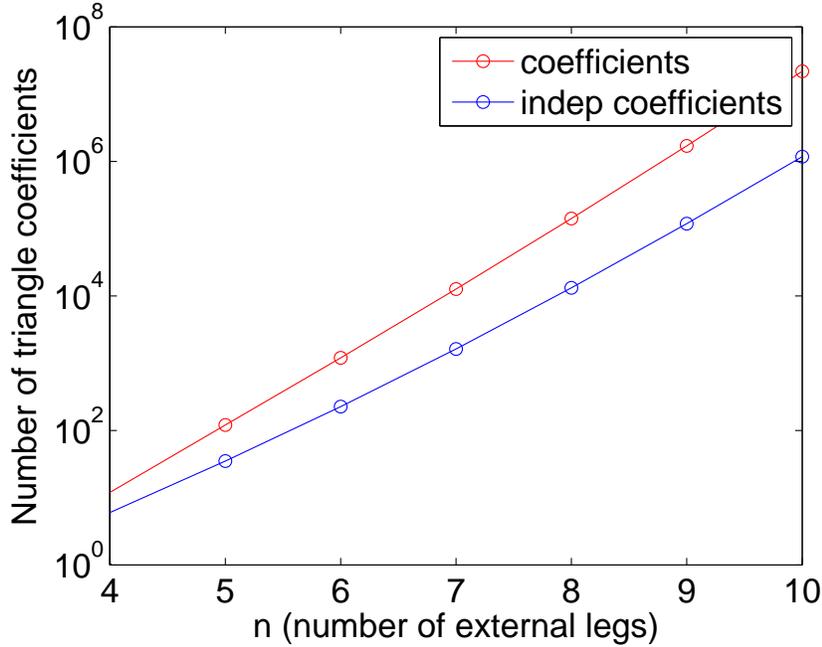}
\caption{The number of triangle coefficients. The red line represents the number of independent coefficients before using the BCJ integral coefficient relations, and the blue line represents the number of independent coefficients needed after the relations are taken into consideration.}
\label{plottri}
\end{center}
\end{figure}

In Fig.~\ref{plottri}, we plot the number of triangle coefficients needed before and after the BCJ integral coefficient relations are taken into account. We notice that the number of triangle coefficients is reduced even less than the boxes. This is due to the fact that less cuts puts more legs on a particular tree amplitude, which makes the BCJ relations more plentiful. We found that the absolute value of the Stirling number of the first kind $S_n^3$ gives the correct number of independent $n$-point coefficients. Next, we review the bubble integral coefficient identities.

\subsection{BCJ bubble integral coefficient identities}
We now look to see if the BCJ relations can be utilized with the bubbles. Typically, to calculate the bubble coefficient, one has to subtract out the triangle contributions to the bubble coefficient. However, since we already showed that the triangles follow the BCJ coefficient relations for all orders of $t$, we only need to show that the identity is valid for the bubble cut component of the bubble integral coefficient. 

In particular, we analytically and numerically checked that the solution works for the $(41,23)$ cut for the amplitude $A^\textrm{one-loop}(1^-,2^-,3^+,4^+)$. For example, we would like to see how the coefficient $b_{(41,32)}$ of $A^\textrm{one-loop}(1^-,3^+,2^-,4^+)$ could be found from $b_{(41,23)}$ using the BCJ relations,  
\begin{equation}
b_{(12,43)}^\pm(y,t) =  \frac{s_{l4}(t,y)}{s_{-l_14}(t,y)}b^\pm_{(12,34)}(t,y).
\end{equation}
Note that to find the true bubble coefficient, we must properly remove the $y$ and $t$ dependence, such that
\begin{equation}
b_{(12,43)}^\pm = \left.\left[\mbox{Inf}_t\left[\mbox{Inf}_y\frac{s_{l4}(t,y)}{s_{-l_14}(t,y)}b^\pm_{(12,34)}(t,y)\right]\right]\right|_{t=0,y^i=Y_i}.
\label{eqn:55}
\end{equation}

\begin{figure}
\begin{center}
\includegraphics[scale=.8]{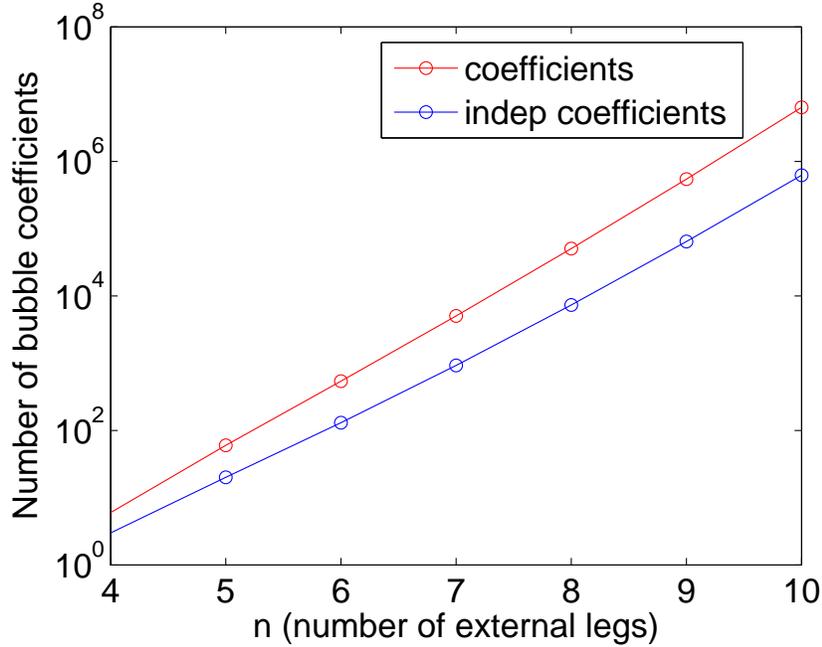}
\caption{The number of bubble coefficients.
We did not include one-mass coefficients, since the corresponding integrals integrate to zero.}
\label{plotbub}
\end{center}
\end{figure}

In Fig.~\ref{plotbub}, we plot the number of bubble coefficients needed before and after the BCJ integral coefficient relations are taken into consideration. As expected, the bubbles are simplified even more heavily than the triangles and boxes. Similar to the triangles, we can use the Stirling number of the first kind $S_n^2$ to count the number of independent coefficients, but this also includes one-mass coefficients. To find the numbers shown in Fig.~\ref{plotbub}, we subtracted $n(n-2)!$.

We have clearly demonstrated that the tree-level BCJ amplitude relations can be used to create one-loop integral coefficient relations. However, we note that these BCJ integral coefficient relations are only useful for amplitudes with multiple identical particles, which often is the case for QCD jet processes. However, there will always be other particles interacting with these gluons, which would lessen the number of identities which are suggested by the figures shown throughout this section. These relations could be useful for improving the efficiency of QCD calculations or to check the stability of numerical code.

\section{Examples of BCJ integral coefficient relations}
\label{sec:4}

\subsection{Box integral coefficient relation example}
Let us consider the box integral coefficients $d_{(1,2,34,5)}$ and $d_{(1,2,43,5)}$ and show explicitly that these coefficients satisfy the integral coefficient relation provided in this work. We start by calculating the $d_{(1,2,34,5)}$ diagram explicitly, shown in Fig.~\ref{fig4}. In principle, there are eight possible loop helicity configurations, but only one is non-zero for this specific cut. We can write down the coefficient by multiplying by the four tree amplitudes:
\begin{figure}
\begin{center}
\includegraphics[scale=.6]{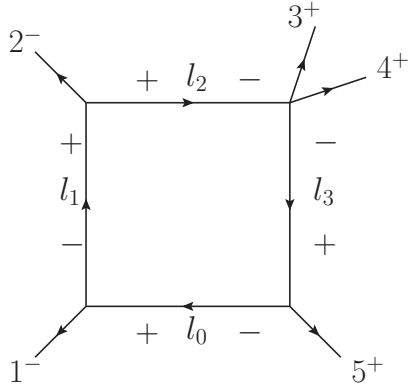}
\caption{The one non-zero loop helicity configuration is shown above for the coefficient $d_{(1,2,34,5)}$.}
\label{fig4}
\end{center}
\end{figure}
\begin{equation}
d_{(1,2,34,5)} = \frac{i\aad{1l_1}^3}{\aad{l_1\!-\!l}\aad{-l1}}\frac{-i\ssd{l_2\!-\!l_1}^3}{\ssd{-l_12}\ssd{2l_2}}\frac{i\aad{l_3\!-\!l_2}^3}{\aad{-l_23}\aad{34}\aad{4l_3}}\frac{-i\ssd{-l_35}^3}{\ssd{5l}\ssd{l\!-\!l_3}} = \frac{\asf{1}{l_1l_2l_3}{5}^3}{\aad{34}\asf{1}{l}{5}\asf{3}{l_2}{2}\asf{4}{l_3ll_1}{2}},
\end{equation}
where $l=l_0$ throughout. We can use the loop solution from Eq.~(\ref{eqn:18}) and find $l$ to be
\begin{eqnarray}
l^+ &=& \frac{1}{2}\frac{\ssd{21}}{\ssd{25}}\asf{1}{\gamma^\mu}{5}, \nonumber\\
l^- &=& \frac{1}{2}\frac{\aad{21}}{\aad{25}}\asf{5}{\gamma^\mu}{1}.
\end{eqnarray}
Right away, it is clear that the positive solution gives zero, since $\aad{1l^+}=0$. We continue by only considering the negative loop solution. We can make a choice for $\left<l^-\right|$ and $|l^-]$, 
\begin{equation}
\left<l^-\right| = \left<5\right|,\textrm{ }\textrm{ }\textrm{ }\textrm{ }\textrm{ }|l^-] = \frac{\aad{21}}{\aad{25}}|1]\equiv \alpha|1],
\end{equation}
which allows us to simplify the calculation in terms of external spinors. 

Futhermore, we can use momentum conservation to express all of the other loop momenta in terms of $l$:
\begin{eqnarray}
d^-_{(1,2,34,5)} &=& \frac{\asf{1}{l}{2}^3\asf{2}{l}{5}^3}{\aad{34}\asf{1}{l}{5}\asf{3}{l-1}{2}\aad{45}\asf{1}{l}{5}\ssd{12}} \nonumber\\
&=& \frac{\aad{15}^3\alpha^3\ssd{12}^3\aad{25}^3\alpha^3\ssd{15}^3}{\aad{34}\aad{45}(\aad{15}\alpha\ssd{15})^2(\aad{35}\alpha\ssd{12}-\aad{31}\ssd{12})\ssd{12}} \nonumber \\
&=& is_{51}s_{12}\frac{i\aad{12}^3}{\aad{23}\aad{34}\aad{45}\aad{51}} = is_{51}s_{12}A_5^{\textrm{tree}}(1^-,2^-,3^+,4^+,5^+).
\end{eqnarray}
In the last line, we used the Schouten identity to simplify the denominator. Similarly, we can immediately write down the equation for the box coefficient $c_{(1,2,43,5)}$ since $p_3$ and $p_4$ have the same helicity, which is given by
\begin{equation}
d^-_{(1,2,43,5)} =  is_{51}s_{12}\frac{i\aad{12}}{\aad{24}\aad{43}\aad{35}\aad{51}}= is_{51}s_{12}A_5^{\textrm{tree}}(1^-,2^-,4^+,3^+,5^+).
\end{equation}
Next, we check that the coefficient cut relation
$d^-_{(1,2,43,5)}=\frac{s_{l^-_34}}{s_{-l^-_24}}d^-_{(1,2,34,5)}$
holds. To start,
\begin{eqnarray}
\frac{s_{l_3^-4}}{s_{-l_2^-4}} &=& \frac{\asf{4}{l_3}{4}}{\asf{4}{-l_2}{4}} = \frac{\asf{4}{l+5}{4}}{\asf{4}{1+2-l}{4}},\nonumber\\
\frac{s_{l^-_34}}{s_{-l^-_24}} &=& \frac{-\frac{\aad{45}}{\aad{25}}(\asf{2}{1+5}{4})}{\frac{\ssd{14}}{\aad{25}}(\aad{45}\aad{21}-\aad{41}\aad{25})-\asf{4}{2}{4}} = -\frac{\aad{45}\aad{23}}{\aad{24}\aad{35}}.
\end{eqnarray}
Schouten identities and momentum conservation are used throughout to simplify these expressions. We can see that this is the exact factor which is needed to find $d^-_{(1,2,43,5)}$ from $d^-_{(1,2,34,5)}$, since
\begin{eqnarray}
d_{(1,2,43,5)}^+ &=& \frac{s_{l^+_34}}{s_{-l^+_24}}d_{(1,2,34,5)}^+= 0, \nonumber\\
d_{(1,2,43,5)}^- &=& \frac{s_{l^-_34}}{s_{-l^-_24}}d_{(1,2,34,5)}^- = is_{51}s_{12}\frac{i\aad{12}}{\aad{24}\aad{43}\aad{35}\aad{51}}.
\end{eqnarray}
As we have demonstrated, the BCJ integral coefficient identity holds for this five-point box cut example.

\subsection{Triangle integral coefficient relation example}
Next, we will show how the BCJ integral coefficient relation holds for a triangle cut. We will choose a four-point cut $c_{(1,23,4)}$ which has a zero triangle coefficient, but does have non-zero terms for powers of $t$ greater than zero. These higher power terms contribute to the bubble coefficient, so we must confirm this in order to show that the BCJ integral coefficient relations work on the total bubble coefficient. In this example, the BCJ relations will be used to find a non-zero triangle integral coefficient from a zero triangle integral coefficient, which is possible since the inverse propagator ratio has t dependence. 
\begin{figure}
\begin{center}
\includegraphics[scale=.6]{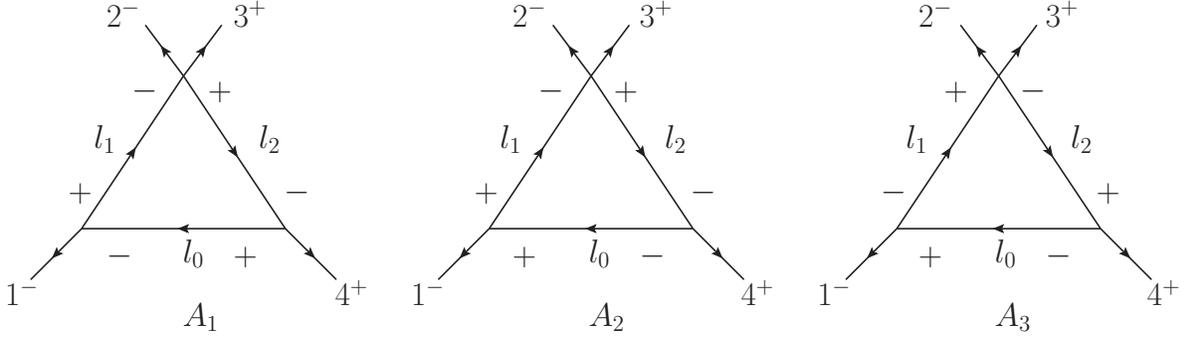}
\caption{Diagrams needed for the coefficient $c_{(1,23,4)}$.}
\label{fig5}
\end{center}
\end{figure}

We will calculate the triangle cut from the amplitude
$A_4^{\textrm{one-loop}}(1^-,2^-,3^+,4^+)$. There are three nonzero
loop helicities which we must consider. We have the three diagrams,
which we will label as $A_1$, $A_2$, and $A_3$ and are shown in
Fig.~\ref{fig5}, which when evaluated gives
\begin{eqnarray}
A_1 &=& \frac{-i\aad{l1}^3}{\aad{1l_1}\aad{l_1l}}\frac{\aad{l_12}^3}{\aad{23}\aad{3l_2}\aad{l_2l_1}}\frac{\ssd{4l}^3}{\ssd{ll_2}\ssd{l_24}}, \nonumber\\
A_2 &=& \frac{-i\ssd{l_1l}^3}{\ssd{l1}\ssd{1l_1}}\frac{\aad{l_12}^3}{\aad{23}\aad{3l_2}\aad{l_2l_1}}\frac{\aad{ll_2}^3}{\aad{l_24}\aad{4l}}, \nonumber\\
A_3 &=& \frac{-i\aad{1l_1}^3}{\aad{l_1l}\aad{l1}}\frac{\aad{2l_2}^4}{\aad{23}\aad{3l_2}\aad{l_2l_1}\aad{l_12}}\frac{\ssd{l_24}^3}{\ssd{4l}\ssd{ll_2}}.
\end{eqnarray}
Choosing $K_1^{\flat,\mu} = p_1^\mu$ and $K_3^{\flat,\mu} = p_4^\mu$, we find that the following loop solution is
\begin{eqnarray}
&\left<l^+\right| = t\left<1\right|,  &\mbox{ }\mbox{ }\mbox{ }\mbox{ }\mbox{ } |l^+] = |4], \nonumber\\
&\left<l_1^+\right| = t\left<1\right|,  &\mbox{ }\mbox{ }\mbox{ }\mbox{ }\mbox{ } |l_1^+] = |4] - \frac{1}{t}|1], \nonumber \\
&\left<l_2^+\right| = t\left<1\right| + \left<4\right|,  &\mbox{ }\mbox{ }\mbox{ }\mbox{ }\mbox{ } |l_2^+] = |4].
\end{eqnarray}
This leaves three non-zero contributions after considering the positive and negative loop solutions, 
\begin{eqnarray}
A_2^+ &=& \frac{i\ssd{41}\aad{12}^3}{\aad{23}(t\aad{31}+\aad{34})}, \nonumber \\
A_1^- &=& \frac{-i(\aad{42}-\frac{1}{t}\aad{12})^3t^3\ssd{41}}{\aad{23}\aad{34}}, \nonumber \\
A_3^- &=& \frac{i\aad{24}^4t^3\ssd{14}}{\aad{23}\aad{34}(\aad{42}-\frac{1}{t}\aad{12})}.
\end{eqnarray}
After expanding about $t=\infty$, we find that only the negative
solution has non-zero contributions to the triangle and bubble
coefficients. Next, we can calculate $c_{(1,23,4)}(t)$, which is the
same as $c_{(1,23,4)}$ before picking the $t^0$ term after the
expansion about infinity. We find
\begin{eqnarray}
c^+_{(1,23,4)}(t) &=& 0 + \mathcal{O}\left(\frac{1}{t}\right), \nonumber\\
c^-_{(1,23,4)}(t) &=& \frac{2i\aad{24}\ssd{41}}{\aad{23}\aad{34}}(\aad{24}^2t^3 + \aad{24}\aad{12}t^2 + 2\aad{12}^2t) + \mathcal{O}\left(\frac{1}{t}\right).
\end{eqnarray}
The fact that there is no zeroth order term shows that there is no triangle coefficient, yet the higher powers of $t$ would feed into the bubble coefficient. Let us check the BCJ integral coefficient relation by first calculating the coefficient $c^\pm_{(1,32,4)}(t)$ and then confirming that the two coefficients satisfy the corresponding BCJ relation.

\begin{figure}
\begin{center}
\includegraphics[scale=.6]{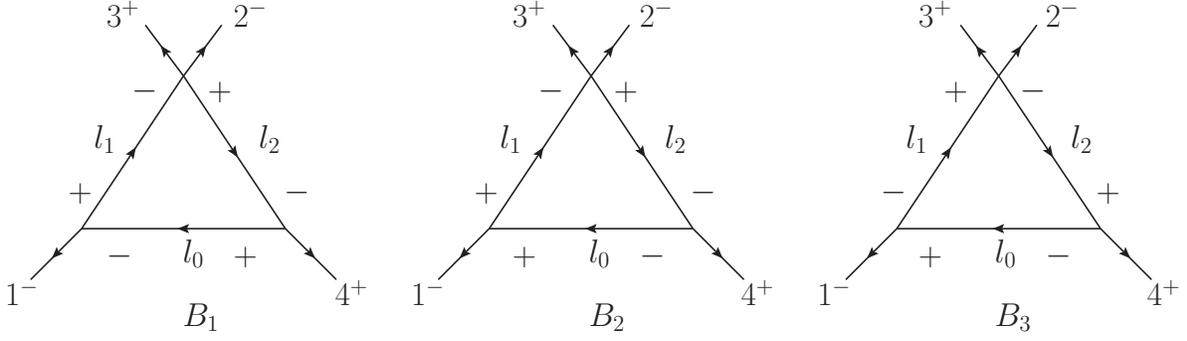}
\caption{Diagrams needed for the coefficient $c_{(1,32,4)}$.}
\label{fig6}
\end{center}
\end{figure}

Once again, we can write down expressions for the three amplitudes, which we will refer to as $B_1$, $B_2$, and $B_3$ and are shown in Figure \ref{fig6}.

\begin{eqnarray}
B_1 &=& \frac{-i\aad{l1}^3}{\aad{1l_1}\aad{l_1l}}\frac{\aad{2l_1}^4}{\aad{2l_2}\aad{l_2l_1}\aad{l_13}\aad{32}}\frac{\ssd{4l}^3}{\ssd{ll_2}\ssd{l_24}}, \nonumber\\
B_2 &=& \frac{-i\ssd{l_1l}^3}{\ssd{l1}\ssd{1l_1}}\frac{\aad{2l_1}^4}{\aad{2l_2}\aad{l_2l_1}\aad{l_13}\aad{32}}\frac{\aad{ll_2}^3}{\aad{l_24}\aad{4l}}, \nonumber \\
B_3 &=& \frac{-i\aad{1l_1}^3}{\aad{l_1l}\aad{l1}}\frac{\aad{2l_2}^3}{\aad{l_2l_1}\aad{l_13}\aad{32}}\frac{\ssd{l_24}^3}{\ssd{4l}\ssd{ll_2}}.
\end{eqnarray}
Similarly, there are three non-zero contributions to the triangle/bubble integral coefficient, which are shown below. We used the same loop solution as previously and find
\begin{eqnarray}
B_2^+ &=& \frac{i\ssd{41}\aad{12}^4}{\aad{13}\aad{32}(t\aad{21}+\aad{24})}, \nonumber \\
B_1^- &=& \frac{-i\left(\aad{24}-\frac{1}{t}\aad{21}\right)^4t^3\ssd{41}}{\aad{24}\aad{32}\left(\aad{43}-\frac{1}{t}\aad{13}\right)}, \nonumber \\
B_3^- &=& \frac{it^3\aad{24}^3\ssd{14}}{\aad{32}\left(\aad{43}-\frac{1}{t}\aad{13}\right)}.
\end{eqnarray}
Once again, we find that the positive solution has no contribution to the bubble or triangle coefficient. Interestingly enough, the negative solution does have a non-zero triangle integral coefficient. Since the analytic expression for this amplitude is a bit lengthy, we will only report the zeroth order term, which corresponds to the triangle integral coefficient.
\begin{eqnarray}
c^+_{(1,32,4)} &=& 0, \nonumber \\
c^-_{(1,32,4)} &=& \frac{2is_{41}(\aad{13}\aad{24}\aad{14}\aad{23} + 2\aad{12}^2\aad{34}^2)}{\aad{34}^4}.
\end{eqnarray}
Next, we show that we get the same result if we were to use the BCJ integral coefficient relations. The ratio of inverse propagators $\frac{s_{l_23}}{s_{-l_13}}$ for the positive and negative loop solution are
\begin{eqnarray}
\frac{s_{l^+_23}}{s_{-l^+_13}} &=& \frac{\ssd{43}(t\aad{13}+\aad{43})}{\aad{13}(t\ssd{34}-\ssd{31})}, \nonumber \\
\frac{s_{l^-_23}}{s_{-l^-_13}} &=& \frac{\aad{34}(t\ssd{31}+\ssd{34})}{\ssd{31}(t\aad{43}-\aad{13})}.
\end{eqnarray}
We see that these two are complex conjugates of each other, if $t$ is real. Next, we will multiply these by $b^\pm_{(1,23,4)}$, expand about $t$ approaches infinity, and keep the zeroth order term. S@M and Mathematica easily allow for this analytic expansion to be performed~\cite{Maitre:2007jq}, which gives
\begin{eqnarray}
 c^+_{(1,32,4)} = \left.\mbox{Inf}_t\left[\frac{s_{l^+_23}(t)}{s_{-l^+_13}(t)} c^+_{(1,23,4)}(t)\right]\right|_{t=0} &=& 0, \nonumber \\
 c^-_{(1,32,4)} = \left.\mbox{Inf}_t\left[\frac{s_{l^-_23}(t)}{s_{-l^-_13}(t)} c^-_{(1,23,4)}(t)\right]\right|_{t=0} &=& \frac{2is_{41}(\aad{13}\aad{24}\aad{14}\aad{23} + 2\aad{12}^2\aad{34}^2)}{\aad{34}^4}.
\end{eqnarray}
After some factoring and application of the Schouten identity, one can get the BCJ integral coefficient relation to give the correct expression for the coefficient $c^\pm_{(1,32,4)}$. Furthermore, we numerically confirmed that the coefficients agree for all orders of $t$, not just for the $t^0$ term. This ensures that the triangle contributions to the bubbles will also satisfy the BCJ integral coefficient relations.

\subsection{Bubble integral coefficient relation example}
\begin{figure}
\begin{center}
\includegraphics[scale=.6]{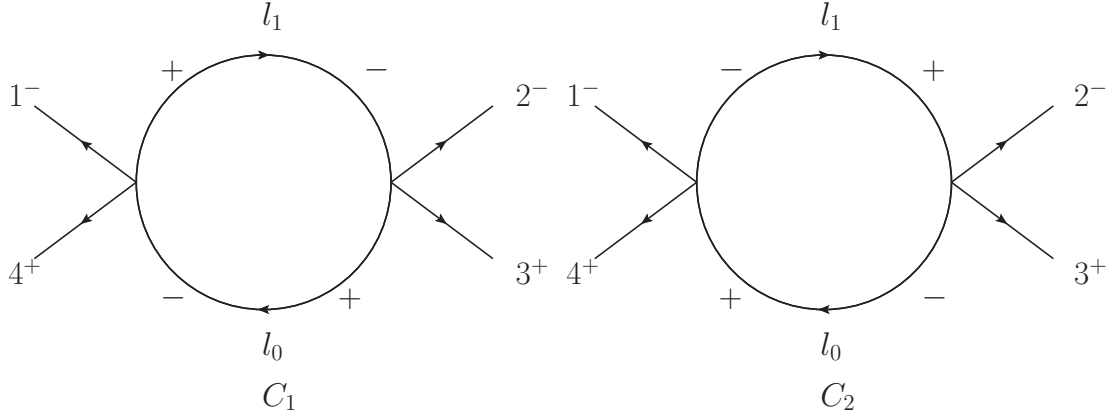}
\caption{The non-zero helicity configurations for the coefficient $b_{(41,23)}$.}
\label{fig7}
\end{center}
\end{figure}

In this subsection, we present a calculation of a four-point bubble coefficient and show that it satisfies a BCJ integral coefficient relation. We start by considering the $b_{(41,23)}$ integral coefficient of the amplitude $A_4^{\textrm{one-loop}}(1^-,2^-,3^+,4^+)$, which is shown in Figure \ref{fig7}.

We have two non-zero internal helicity configurations to consider, which we will refer to as $C_1$ and $C_2$. 
\begin{eqnarray}
C_1 &=& \frac{i\aad{1l}^4}{\aad{1l_1}\aad{l_1l}\aad{l4}\aad{41}}\frac{i\aad{l_12}^3}{\aad{23}\aad{3l}\aad{ll_1}}, \nonumber \\
C_2 &=& \frac{i\aad{1l_1}^3}{\aad{l_1l}\aad{l4}\aad{41}}\frac{i\aad{2l}^4}{\aad{23}\aad{3l}\aad{ll_1}\aad{l_12}}.
\end{eqnarray}
Each also has two loop solutions, giving $C^\pm_1$ and $C^\pm_2$. We chose $\chi^\mu = p_1^\mu$, which makes $K_1^{\flat,\mu} = p_4^\mu$. The loop solutions are
\begin{eqnarray}
\left<l^+\right| = t\left<4\right|+(1-y)\left<1\right|,&\mbox{ }\mbox{ }\mbox{ }\mbox{ }\mbox{ }& |l^+] = \frac{y}{t}|4]+|1], \nonumber \\
\left<l_1^+\right| = \left<4\right|-\frac{y}{t}\left<1\right|,&\mbox{ }\mbox{ }\mbox{ }\mbox{ }\mbox{ }& |l_1^+] = (y-1)|4]+t|1].
\end{eqnarray}
Plugging these solutions in and simplifying, we find
\begin{eqnarray}
C^+_1 &=& \frac{t(t\aad{42}-y\aad{12})^3}{\aad{23}\aad{41}(1-y)(t\aad{34}+(1-y)\aad{31})}, \nonumber \\
C^+_2 &=& \frac{t(t\aad{24}+(1-y)\aad{21})^4}{\aad{23}\aad{41}(1-y)(t\aad{34}+(1-y)\aad{31})(t\aad{42}-y\aad{12})}, \nonumber \\
C^-_1 &=& \frac{\left(\frac{y}{t}\right)^4((y-1)\aad{42}+t\aad{12})^3}{\aad{23}\aad{41}(y-1)\left(\frac{y}{t}\aad{34}+\aad{31}\right)}, \nonumber \\
C^-_2 &=& \frac{(y-1)^3\left(\frac{y}{t}\aad{24}+\aad{21}\right)^4}{\aad{23}\aad{41}\left(\frac{y}{t}\aad{34}+\aad{31}\right)((y-1)\aad{42}+t\aad{12})}.
\end{eqnarray}
To calculate the bubble coefficient, one must typically subtract away the corresponding triangle contributions. However, we have already shown that the triangle contributions will cancel at all orders of $t$, not just the component contributing to the triangle coefficient. Therefore, to confirm that the BCJ integral coefficient relation holds for the bubble coefficient, we will just focus on the bubble cut contribution to the bubble coefficient. 

\begin{figure}
\begin{center}
\includegraphics[scale=.6]{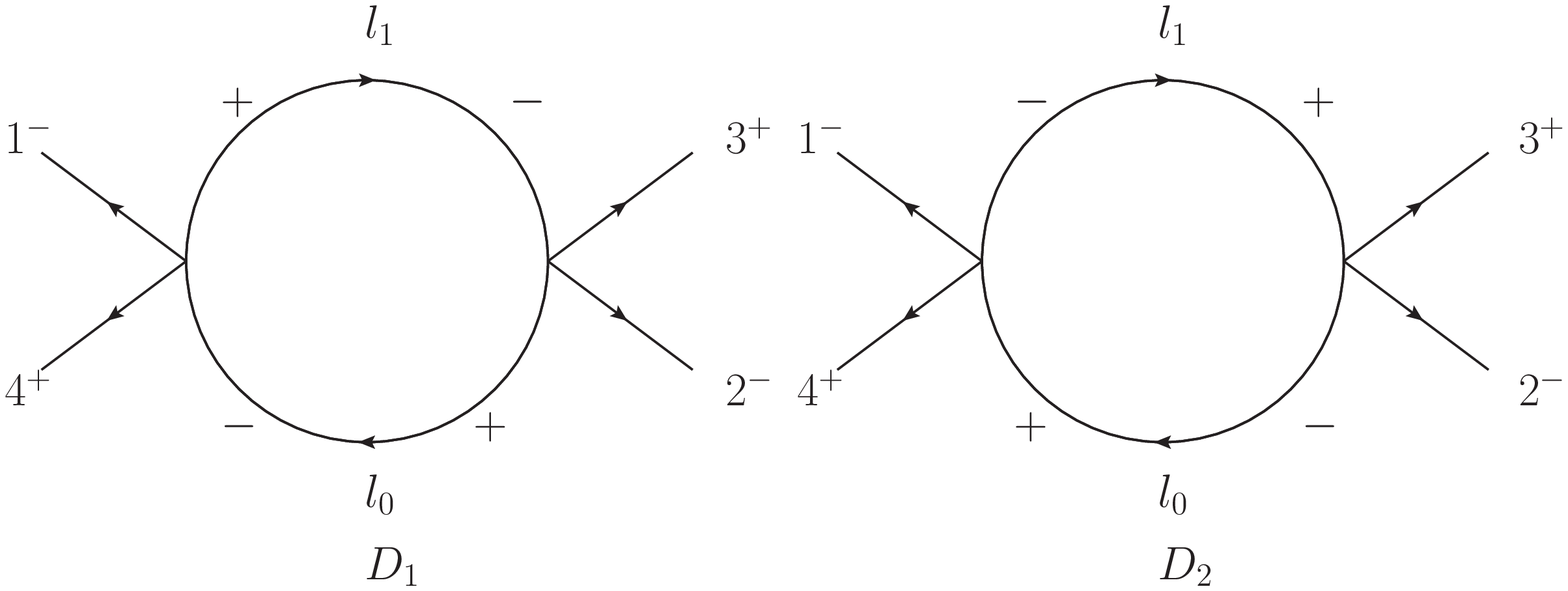}
\caption{The non-zero helicity configurations for the coefficient $b_{(41,32)}$.}
\label{fig8}
\end{center}
\end{figure}

We find the coefficient $b^+_{(41,23)}$ is zero by applying Eq.~(\ref{eqn:36}) to $C^+_1+C^+_2$. For $b^-_{(41,23)}$, we get
\begin{equation}
b^-_{(41,23)} = \frac{2\aad{13}^2\aad{24}^2-\aad{12}\aad{13}\aad{24}\aad{34}+11\aad{12}^2\aad{34}^2}{3\aad{34}^4}.
\end{equation}
Next, we would like to calculate $b_{(41,32)}^\pm$ and see if it can be found from $b^\pm_{(41,23)}$. The former coefficient has four contributions $D^\pm_1$ and $D^\pm_2$, which is shown in Figure \ref{fig8}.

\begin{eqnarray}
D_1 &=& \frac{i\aad{1l}^4}{\aad{1l_1}\aad{l_1l}\aad{l4}\aad{41}}\frac{i\aad{2l_1}^4}{\aad{2l}\aad{ll_1}\aad{l_13}\aad{32}},\nonumber \\
D_2 &=& \frac{i\aad{1l_1}^3}{\aad{l_1l}\aad{l4}\aad{41}}\frac{i\aad{2l}^3}{\aad{ll_1}\aad{l_13}\aad{32}}.
\end{eqnarray}
We can use the same loop solution as before and evaluate $D_\pm^1$ and $D_\pm^2$ to find
\begin{eqnarray}
D_1^+ &=& \frac{t^4\left(\aad{24}-\frac{y}{t}\aad{21}\right)^4}{\aad{32}\aad{41}(1-y)(t\aad{24}+(1-y)\aad{21})\left(\aad{43}-\frac{y}{t}\aad{13}\right)}, \nonumber \\
D_2^+ &=& \frac{(t\aad{24}+(1-y)\aad{21})^3}{\aad{32}\aad{41}(1-y)\left(\aad{43}-\frac{y}{t}\aad{13}\right)}, \nonumber \\
D_1^- &=& \frac{\left(\frac{y}{t}\right)^4((y-1)\aad{24}+t\aad{21})^4}{\aad{32}\aad{41}(y-1)\left(\frac{y}{t}\aad{24}+\aad{21}\right)((y-1)\aad{43}+t\aad{13})}, \nonumber \\
D_2^- &=& \frac{(y-1)^3\left(\frac{y}{t}\aad{24}+\aad{21}\right)^3}{\aad{32}\aad{41}((y-1)\aad{43}+t\aad{13})}.
\end{eqnarray}
Finally, we can use Eq.~(\ref{eqn:36}) and find the contribution to the bubble coefficient $b_{(41,32)}^\pm$. It is no surprise that $b_{(41,32)}^+$ is zero, and we find 
\begin{equation}
b_{(41,32)}^- = \frac{-11\aad{13}^2\aad{24}^2+13\aad{12}\aad{13}\aad{24}\aad{34}-14\aad{12}^2\aad{34}^2}{3\aad{34}^4}.
\end{equation}

Next, we would like to calculate $b_{(41,32)}^\pm$ from $b_{(41,23)}^\pm$ by using the BCJ integral coefficient relation Eq.~(\ref{eqn:55}) and confirm that we get the correct result. First, we find the needed ratio of inverse propagators, which are
\begin{eqnarray}
\frac{s_{l^+3}}{s_{-l_1^+3}} &=& -\frac{(t\aad{43}+(1-y)\aad{13})\left(\frac{y}{t}\ssd{34}+\ssd{31}\right)}{\left(\aad{43}-\frac{y}{t}\aad{13}\right)((y-1)\ssd{34}+t\ssd{31})}, \nonumber \\
\frac{s_{l^-3}}{s_{-l_1^-3}} &=& -\frac{\left(\frac{y}{t}\aad{43}+\aad{13}\right)(t\ssd{34}+(1-y)\ssd{31})}{((y-1)\aad{43}+t\aad{13})\left(\ssd{34}-\frac{y}{t}\ssd{31}\right)}.
\end{eqnarray}
We can apply the BCJ integral coefficient relation to confirm that we get the right result.
\begin{eqnarray}
b^\pm_{(41,32)} &=& \left.\mbox{Inf}_t\left[\mbox{Inf}_y\left[\frac{s_{l^\pm3}(t,y)}{s_{l^\pm_13}(t,y)}b^\pm_{(41,23)}(t,y)\right]\right]\right|_{t=0,y^i=Y_i}, \nonumber \\
&=&  \frac{-11\aad{13}^2\aad{24}^2+13\aad{12}\aad{13}\aad{24}\aad{34}-14\aad{12}^2\aad{34}^2}{3\aad{34}^4}.
\end{eqnarray}
We confirmed that the two solutions agree numerically, thus showing that the BCJ integral coefficients work on bubble coefficients as well.

\section{Conclusions}
\label{sec:5}

The unitarity method implies that tree-level properties can many times
be carried over to loop level.  In this paper we demonstrated that
tree-level BCJ amplitude relations can be recycled into relations
between integral coefficients at loop level.  The relations are
actually not between full coefficients, but rather they are satisfied
separately by the two independent loop-momentum cut solutions used to
construct the integral coefficient. Both solutions are needed when
constructing the coefficient using the unitarity method, so both are
anyway both available. These identities on integral coefficients can
be used to reduce the number of tree amplitudes and cut coefficients
that are needed to find the full one-loop amplitude. Alternatively,
these relations could be used in a numerical code to confirm the
stability of the coefficients.

Future work could include investigating how these relations can be
used to help deal with the rational parts of QCD amplitudes, which are
the most time-exhaustive part of the one-loop amplitude.  It 
would also be interesting to understand the higher-loop implications.

The author would like to thank Zvi Bern, Scott Davies, and Josh Nohle
for many discussions.

\bibliography{coeff}

\end{document}